\documentclass{article}

\usepackage{PRIMEarxiv}

\usepackage[utf8]{inputenc} 
\usepackage[T1]{fontenc}    
\usepackage{hyperref}       
\usepackage{url}   
\usepackage{amsmath}
\usepackage{booktabs}       
\usepackage{amsfonts}       
\usepackage{nicefrac}       
\usepackage{microtype}      
\usepackage{lipsum}
\usepackage{graphicx}
\graphicspath{{media/}}     
\usepackage{amsmath}
\usepackage{graphicx}
\usepackage{geometry}
\usepackage{authblk}

\title{ResNCT: A Deep Learning Model for the Synthesis of Nephrographic Phase Images in CT Urography}
\author[1]{Syed Jamal Safdar Gardezi, PhD}
\author[1]{Lucas Aronson}
\author[2]{Peter Wawrzyn}
\author[2]{Hongkun Yu, MS}
\author[3]{E. Jason Abel, MD}
\author[3]{Daniel D. Shapiro, MD}
\author[1]{Meghan G. Lubner, MD}
\author[1]{Joshua Warner, MD, PhD}
\author[1]{Giuseppe Toia, MD, MS}
\author[4]{Lu Mao, PhD}
\author[1,2]{Pallavi Tiwari, PhD}
\author[1,2,5]{Andrew L. Wentland, MD, PhD\thanks{Corresponding author: Andrew L. Wentland, MD, PhD, Departments of Radiology, Medical Physics, \& Biomedical Engineering, University of Wisconsin School of Medicine \& Public Health, 1111 Highland Avenue, Madison, WI, USA 53705. Email: alwentland@wisc.edu}}

\affil[1]{Department of Radiology, University of Wisconsin School of Medicine \& Public Health, Madison, WI, USA}
\affil[2]{Department of Biomedical Engineering, University of Wisconsin – Madison, Madison, WI, USA}
\affil[3]{Department of Urology, University of Wisconsin School of Medicine \& Public Health, Madison, WI, USA}
\affil[4]{Department of Biostatistics, University of Wisconsin School of Medicine \& Public Health, Madison, WI, USA}
\affil[5]{Department of Medical Physics, University of Wisconsin School of Medicine \& Public Health, Madison, WI, USA}

\date{}

\begin{document}

\maketitle



\begin{abstract}
\textbf{Purpose:} To develop and evaluate a transformer-based deep learning model for the synthesis of nephrographic phase images in CT urography (CTU) examinations from the unenhanced and urographic phases.
\textbf{Materials and Methods:} 
This retrospective study was approved by the local Institutional Review Board. A dataset of 119 patients (mean ± SD age, 65 ± 12 years; 75/44 males/females) with three-phase CT urography studies was curated for deep learning model development. The three phases for each patient were aligned with an affine registration algorithm. A custom model, coined Residual transformer model for Nephrographic phase CT image synthesis (ResNCT), was developed and implemented with paired inputs of non-contrast and urographic sets of images trained to produce the nephrographic phase images, that were compared with the corresponding ground truth nephrographic phase images. The synthesized images were evaluated with multiple performance metrics, including peak signal to noise ratio (PSNR), structural similarity index (SSIM), normalized cross correlation coefficient (NCC), mean absolute error (MAE), and root mean squared error (RMSE).
\textbf{Results:} The ResNCT model successfully generated synthetic nephrographic images from non-contrast and urographic image inputs. With respect to ground truth nephrographic phase images, the images synthesized by the model achieved high PSNR (27.8 ± 2.7 dB), SSIM (0.88 ± 0.05), and NCC (0.98 ± 0.02), and low MAE (0.02 ± 0.005) and RMSE (0.042 ± 0.016).
\textbf{Conclusion:} The ResNCT model synthesized nephrographic phase CT images with high similarity to ground truth images. The ResNCT model provides a means of eliminating the acquisition of the nephrographic phase with a resultant 33\% reduction in radiation dose for CTU examinations.
\end{abstract}


\section{Introduction}
\paragraph{}CT urography (CTU) is commonly performed for the evaluation of hematuria \cite{Silverman2009current}. Hematuria can be caused by stones, infection, renal masses, or urothelial tumors. Initially in a conventional 3-phase CT urography study, non-contrast images are acquired to evaluate for stones and to establish a baseline for potential tumor enhancement on subsequent CT phases. Secondly, a scan is repeated 90-100 seconds after the intravenous administration of iodinated contrast. This temporal delay allows for imaging of the nephrographic phase of the kidney. A third scan is acquired 5\raisebox{0.5ex}{~\texttildelow~}10 minutes after the initial contrast injection, which allows for imaging of the urographic phase. At this final time point, contrast has been excreted into the renal collecting system. Each of these three phases provides unique information about the kidney and collecting systems, and each is invaluable in the workup of hematuria. However, 3-phase CT urography requires approximately three times the radiation dose of a standard portal-venous-phase CT \cite{nawfel2004patient,cheng2019ct,potenta2015ct}.
\paragraph{}As an alternative to the three-phase CTU technique, a split-bolus CT urography technique has previously been developed to reduce radiation dose by combining the nephrographic and urographic phases into a single acquisition \cite{chow2007split}. However, this technique inherently lacks a dedicated nephrographic phase set of images, and moreover prior studies have shown that split-bolus CT urography provides inferior contrast opacification \cite {raman2017upper,morrison2021split}  and reduced distention of the urinary tract \cite{dillman2007comparison} compared to the single bolus three-phase CT urography technique. An ideal alternative method would reduce radiation dose, but preserve the benefits of the single-bolus conventional CT urography technique.
\paragraph{}There is redundancy of information inherent within the nephrographic and urographic phase sets of images, particularly with respect to enhancement of the kidneys. The elimination half-life of iodinated contrast agents is 60\raisebox{0.5ex}{~\texttildelow~}90 minutes in subjects with normal renal function and substantially longer in those with impaired renal function\cite{pasternak2012clinical} . Therefore, substantial enhancement of the renal parenchyma remains on the 10-minute delayed urographic phase set of images. This redundancy of information can be exploited to simplify and optimize the three-phase CTU technique. 
\paragraph{}The purpose of this study is to develop a deep learning model that can synthesize nephrographic phase CT images from the matching non-contrast and urographic phase images. It is hypothesized that a transformer-based model based on ResViT \cite{dalmaz2022resvit} will be able to provide high-quality and accurate synthesized nephrographic phase images from the other CTU phases. The ability to successfully synthesize nephrographic phase images would effectively reduce the CT urography acquisition from a three-phase to a two-phase study, reduce radiation dose by 33\%, provide the dedicated nephrographic phase omitted from the split-bolus CTU method, and eliminate the temporal variation associated with nephrographic phase images.

\section{{Materials and Methods}}
\label{sec:headings}

\paragraph{}A custom Residual transformer model for Nephrographic phase CT image synthesis (ResNCT) based on the previously published residual transformer (ResViT) model \cite{dalmaz2022resvit} was developed for the synthesis of nephrographic phase images from paired inputs of non-contrast and urographic images, as shown in Figure 1. The ResNCT model consists of an encoder, an information bottleneck pathway, and a discriminator subnetwork with a decoder pathway and convolutional layers. The encoder in the ResNCT model consists of a stack of identical layers with a multi-head attention layer and a positional forward feed network. The encoder maps an input sequence of representations that encodes feature maps into the information bottleneck module. The encoder uses multi-head attention and multi-attention fusion transformer (MAFT) blocks to map input representations to encoded feature maps. The primary objective of the information bottleneck module is to refine and improve the encoded features extracted during a specific task. This module accomplishes this by integrating two essential components: local features and global contextual information from the MAFT block. The decoder, based on transposed convolutional layers, synthesizes contrasts regardless of source-target configurations. ResNCT is a unified transformer model that employs these components for image generation and discrimination. In the ResNCT model there are 12 attention heads, and each head processes information separately and their results are combined. This process enables the model to learn different perspective representations of the data.

\begin{figure*}[t!]
\centering
\includegraphics[width=0.9\textwidth]{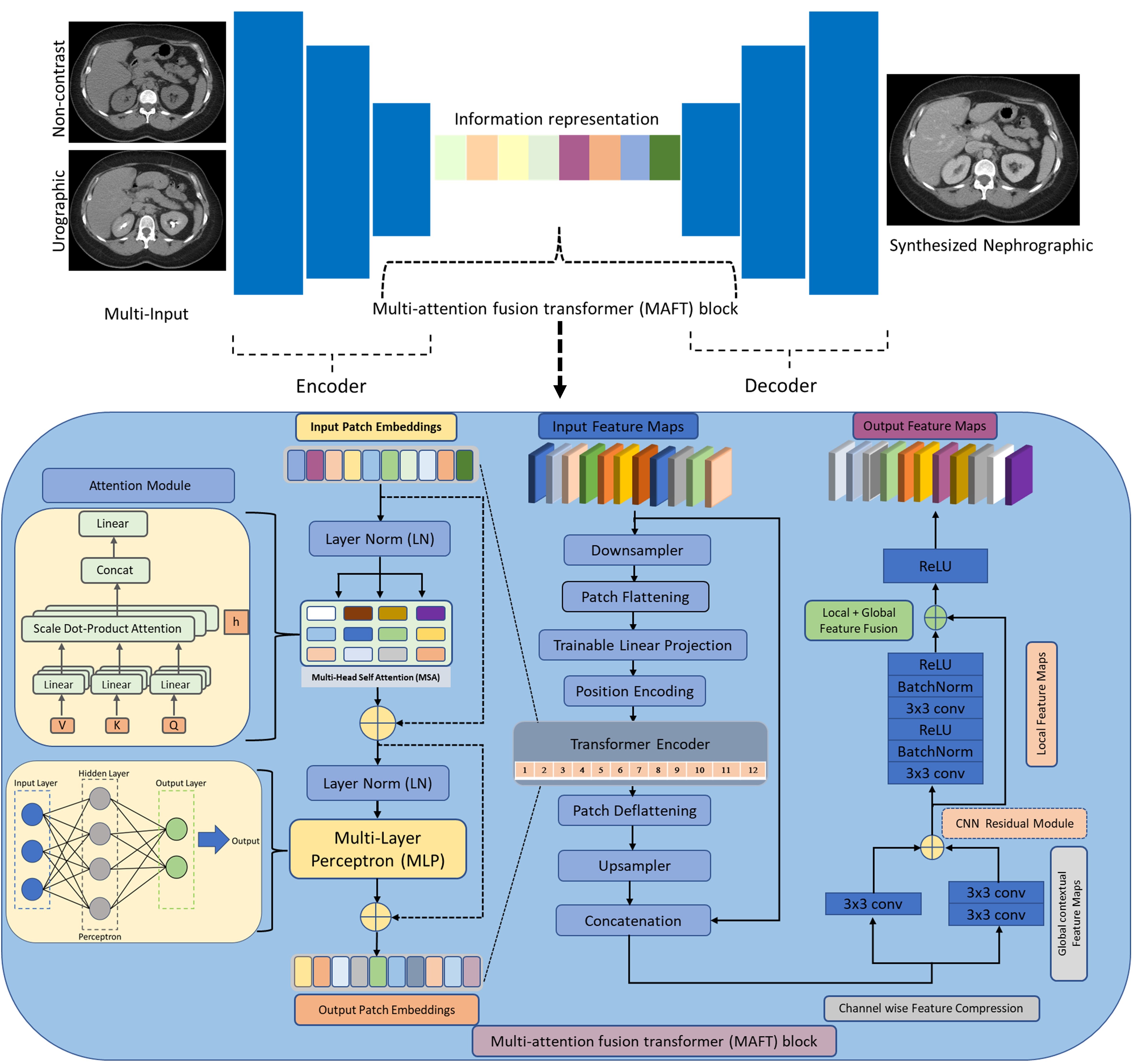}
\caption{\label{fig:fig1.jpg}ResNCT model architecture overview, which utilizes matching non-contrast and urographic phase inputs and outputs a synthesized nephrographic phase image. Encoder-decoder layers are joined by an information block to refine task-specific information. The encoder- decoder layers consist of convolutional operations to learn the structural representations of images. The central information representation block is comprised of cascaded MAFT blocks with a self-attention module to learn global contextual representations with a vision transformer.}
\end{figure*}

\subsection{Dataset}
\paragraph{}This study was approved by the local Institutional Review Board. Informed consent was waived given the retrospective nature of the study. The study included adult patients with three-phase urography, presenting with hematuria, with no known cancer. A dataset of 119 patients (mean ± SD age, 65 ± 12 years; 75/44 males/females) with three-phase CT urography studies was curated for model training. The dataset contains scan acquired on three scanners from three different manufacturer with 90\% from GE Healthcare, 7.5\% from Toshiba and 2.5\% from Siemens systems with slice thickness of
2 - 5 mm, pixel spacing  0.625 - 0.976 mm
 and image resolution of 512 × 512 pixels.
\paragraph{}To reduce the complexity and volume of data for training, a fellowship-trained abdominal radiologist with 7 years of experience selected axial slices that contained only the kidneys from all three CTU phases. Prior to additional processing, images were set to a soft tissue window using a window width of 400 and a window center/level of 50.

\subsection{Image Registration}
\paragraph{}Each of the three phases in a CTU examination are acquired under separate breath holds. Depending on the depth of inspiration, the kidneys can vary in craniocaudal position up to approximately 4 cm \cite{schwartz1994kidney}. Therefore, on a per-patient basis, images were registered using an affine registration algorithm with 12 degrees of freedom. Non-contrast images were used as the fixed image, while the nephrographic and urographic images were used as moving images.
\subsection{Model Training and Performance Metrics}
\paragraph{}Paired inputs of non-contrast and urographic phase images were provided to the ResNCT model, and the matching target nephrographic image was provided for training. Initial training was performed with several learning rates { $2 \times 10^{-3}$, $2 \times 10^{-4}$, $2 \times 10^{-5}$}  and numbers of epochs {200, 500, 700}. 
\paragraph{}An optimal learning rate of $2 \times 10^{-4}$ with  700 epochs was selected for the training set. The model was tuned with 12 attention heads and 3,073 hidden units in each MLP layer. Moreover, since the model uses the discriminator based on Patch-GAN \cite{isola2017image}, the weighting of the pixel-wise, pixel consistency, and adversarial loss respectively were chosen as $\lambda_p = 100$, $\lambda_r = 100$  and  \(\lambda_{\text{adv}} = 100\) respectively. Data on a per-slice basis were split with an 80/20 for training and testing sets using random sampling. The model was developed in a Python 3.8 environment, utilizing PyTorch version 23.07 libraries for its implementation.
\paragraph{}The ResNCT model was compared to existing image-to-image translation methods including CycleGAN, Pix2pix and BicycleGAN \cite{saxena2021comparison} which were similarly trained with an 80/20 split for training and testing sets using identical hyperparameters as the ResNCT model. Several performance metrics were used to compare the ground truth nephrographic phase images to the synthesized nephrographic images from the ResNCT, CycleGAN, Pix2pix, and BicycleGAN models, including peak signal to noise ratio (PSNR), structural similarity index (SSIM), normalized cross correlation coefficient (NCC), mean absolute error (MAE), and root mean squared error (RMSE). Attenuation values from line profiles over the kidneys were obtained from the ground truth nephrographic phase images as well as the synthesized images generated from the ResNCT, CycleGAN, Pix2pix, and BicycleGAN models. Line profiles from the ground truth and synthesized images were compared with RMSE. Furthermore, the attenuation values of cysts and solid masses were compared between ground truth and synthesized nephrographic phase images.
\subsection{Image Quality Analysis}
\paragraph{}For evaluation of image quality, 249 random images including ground truth and synthesized nephrographic phase images from the ResNCT model test set were provided in a blinded fashion to two fellowship- trained abdominal radiologists with 7 and 6 years of experience, respectively. Based on preliminary data, this sample size of \raisebox{0.5ex}{~\texttildelow~}250 images provide a tolerance of 0.1 with 80\% power to detect differences in scores between the ground truth and synthesized images.
\paragraph{}The radiologists were asked to evaluate the image quality, with special focus on the kidneys, using a 4-point Likert scoring system with rating 4 as ‘optimal’, 3 as ‘adequate’, 2 as ‘suboptimal,’ and 1 as ‘Inadequate.’ Inter-rater agreement was assessed by the intraclass correlation coefficient (ICC) for the original score or Cohen’s kappa for the dichotomized score (1-2 vs 3-4). Within each reader, real (ground truth) and synthesized images were compared on the original score using the Wilcoxon rank sum test and on the dichotomized score using the odds ratio (OR) and chi-square test. p values < 0.05 were considered statistically significant.
\section{Results}
\paragraph{}4,200 matched and registered images from the three CTU phases of 119 patients were ultimately utilized for model development. Qualitatively, the ResNCT model was able to provide synthesized nephrographic phase images with high similarity to ground truth images (Figure 2). Synthesized images from the CycleGAN model also successfully provided synthesized nephrographic phase images, albeit with generally more artifacts than the ResNCT model. Synthesized images from the Pix2pix and BicycleGAN models tended to yield the lowest quality images.

\begin{figure*}[t!]
\centering
\includegraphics[width=1\textwidth]{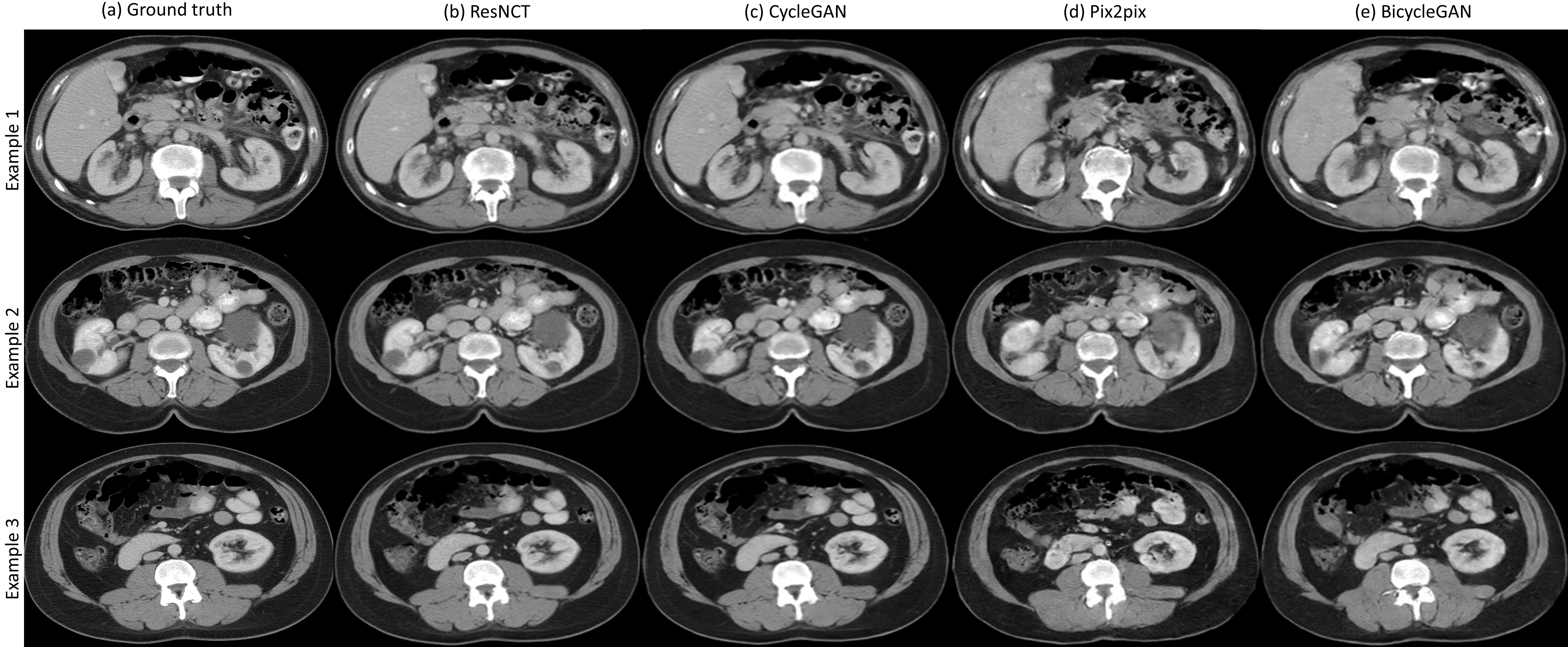}
\caption{\label{fig:fig2.jpg} Example synthetic nephrographic phase images generated by different models. (a) Ground truth (target) image and (b-e) synthesized nephrographic images by the ResNCT, CycleGAN, Pix2pix, and BicycleGAN methods, respectively.}
\end{figure*}

\begin{table}[htbp]
\centering
\caption{Performance metrics comparing four deep learning methods for synthesizing nephrographic phase images in CT urography examinations.}
\label{tab:performance_metrics}
\begin{tabular}{@{}llllll@{}}
\toprule
Methods   & PSNR (dB)          & SSIM               & NCC                & MAE                & RMSE               \\ \midrule
ResNCT    & $27.82 \pm 2.67$   & $0.87 \pm 0.057$   & $0.98 \pm 0.019$   & $0.020 \pm 0.005$  & $0.042 \pm 0.016$  \\
CycleGAN  & $25.22 \pm 0.82$   & $0.93 \pm 0.019$   & $0.99 \pm 0.001$   & $0.028 \pm 0.01$   & $0.055 \pm 0.005$  \\
Pix2pix   & $21.35 \pm 1.39$   & $0.67 \pm 0.06$    & $0.93 \pm 0.026$   & $0.083 \pm 0.02$   & $0.173 \pm 0.030$  \\
BicycleGAN & $17.48 \pm 1.87$  & $0.70 \pm 0.07$    & $0.95 \pm 0.026$   & $0.064 \pm 0.02$   & $0.137 \pm 0.037$  \\ \bottomrule
\end{tabular}

\smallskip

\footnotesize
\begin{flushleft}
\textbf{*} $\pm$ denotes standard deviation.
\end{flushleft}
\end{table}

\subsection{Model Performance}

\paragraph{}The ResNCT model achieved a PSNR of 27.82 ± 2.67 dB, SSIM of 0.87 ± 0.057, NCC of 0.98 ± 0.019, MAE of 0.020 ± 0.005, and RMSE of 0.042 ± 0.016 (Table 1). The CycleGAN yielded a PSNR of 25.22 ± 0.82 dB, SSIM of 0.93 ± 0.019, NCC of 0.99 ± 0.001, MAE of 0.028 ± 0.01, and RMSE of 0.055 ± 0.005 (Table 1). The Pix2pix and BicycleGAN models had broadly inferior performance metrics compared to both the ResNCT and CycleGAN models (Table 1).

\begin{figure*}[t!]
\centering
\includegraphics[width=1\textwidth]{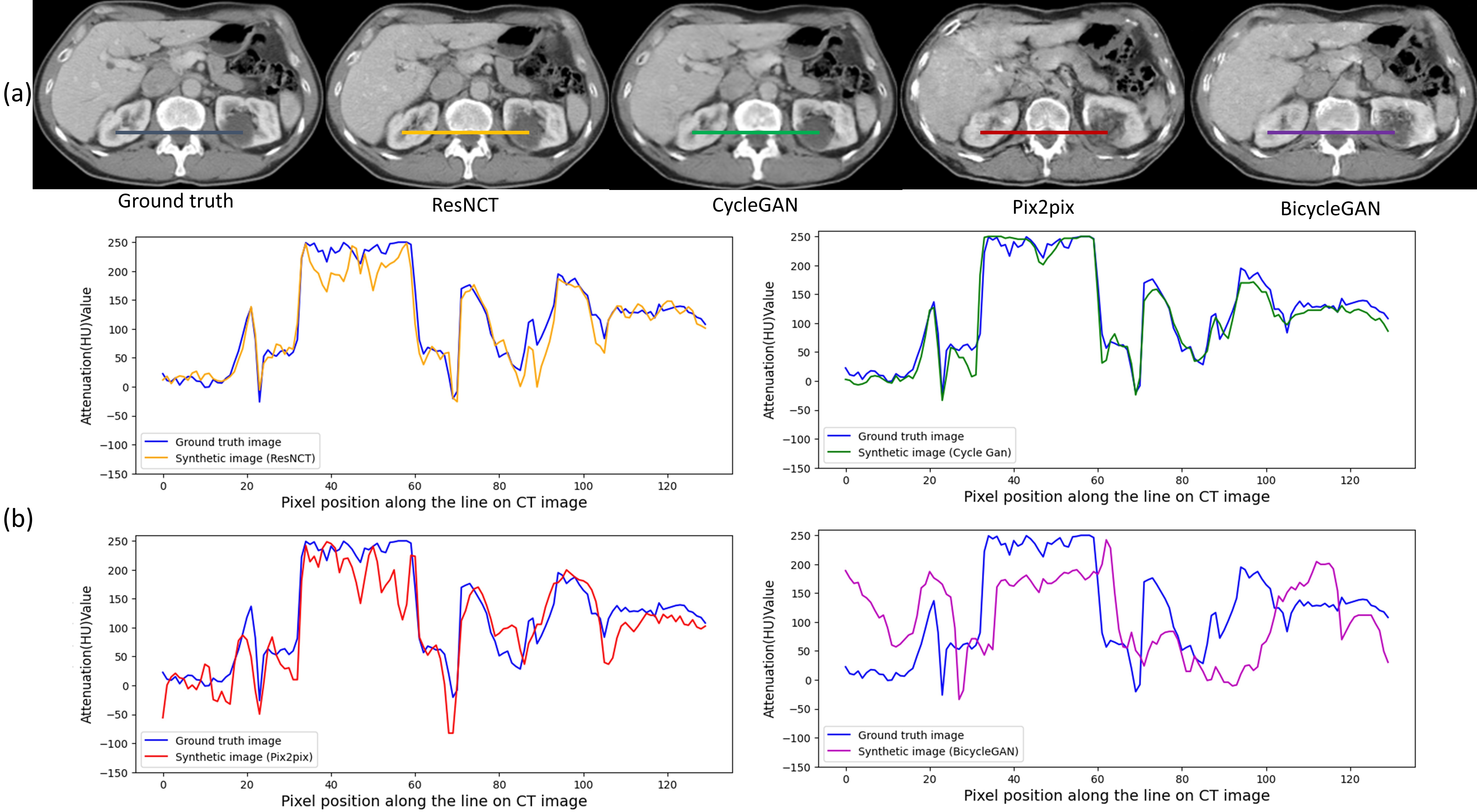}

\caption{\label{fig:fig3.jpg} (a) Ground truth nephrographic phase images as well as synthesized nephrographic phase images from the ResNCT, CycleGAN, Pix2pix, and BicycleGAN models. (b) Attenuation line profiles across the kidneys as shown in part (a) in which attenuation in Hounsfield units is plotted against voxel location along the line profile. Line profiles from ground truth nephrographic phase images are plotted against the line profiles from the synthesized images from the ResNCT, CycleGAN, Pix2pix, and BicycleGAN models.}
\end{figure*}

\paragraph{}Line profiles across the kidneys demonstrated strong agreement between ground truth nephrographic phase images and the synthesized images for the ResNCT model, with a RMSE of 0.0625 (Figure 3).  RMSE was generally higher in comparing the line profiles of the ground truth nephrographic phase images and the synthesized images for the CycleGAN, Pix2pix, and BicycleGAN models, with RMSEs of 0.2278, 0.1038, and 0.0457, respectively (Figure 3).

\paragraph{}The attenuation values within the kidneys were similar between the ground truth and synthesized nephrographic phase images from the ResNCT model (Figure 4), with an overall correlation coefficient of 0.96. A histogram of the attenuation values in the regions of the kidneys were highly similar between the ground truth and synthesized nephrographic phase images from the ResNCT model (Figure 4), with a  RMSE of 0.1546.

\begin{figure*}[t!]
\centering
\includegraphics[height =9.5cm,width=1\textwidth]{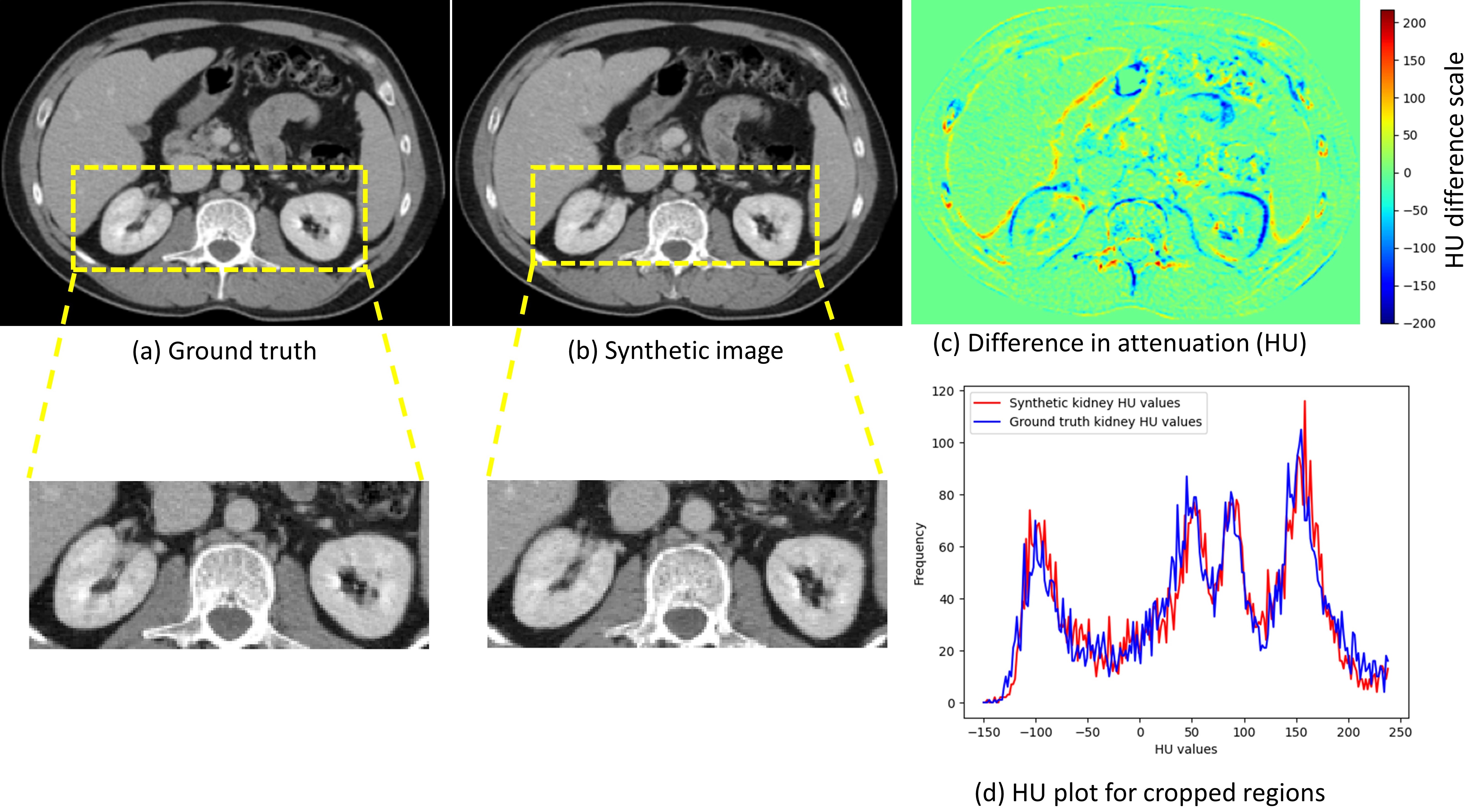}
\caption{\label{fig:fig4.jpg}Analysis of kidney structure on ground truth vs synthetic nephrographic phase images. Attenuation values within the cropped region of the kidneys for ground truth (a) and synthesized (b) nephrographic phase images from the ResNCT model were used to compare attenuation values. A HU difference map (c) for the entirety of the images shows minimal attenuation differences. The region of the kidney specifically shows low differences in a histogram of attenuation values (d).}
\end{figure*}

\paragraph{}Representative renal lesions on synthesized nephrographic phase images from the ResNCT model demonstrate appropriate attenuation values (Figure 5). For example, a solid renal mass with a mean attenuation of 33 HU on non-contrast images enhances to a mean attenuation of 93 HU on ground truth nephrographic phase images, and the synthetic nephrographic phase image from the ResNCT model appropriately has a mean attenuation of 90 HU within the lesion (Figure 5a). Similarly, a cystic renal lesion has a mean attenuation of 11 HU on non-contrast images, does not enhance on ground truth nephrographic phase images with a mean attenuation of 15 HU, and appropriately has a mean attenuation of 17 HU on the synthetic nephrographic phase image from the ResNCT model (Figure 5b).

\begin{table}[htbp]
\centering
\caption{Likert image quality scoring by two fellowship-trained abdominal radiologists for ground truth and synthetic nephrographic phase images from the ResNCT model. The 4-point Likert scoring system used rating 4 as ‘optimal’, 3 as ‘adequate’, 2 as ‘suboptimal,’ and 1 as ‘Inadequate.’}
\label{tab:likert_scoring}
\begin{tabular}{@{}llccccc@{}}
\toprule
Type & Reader(s) & 1 & 2 & 3 & 4 & Overall \\ \midrule
Real & Reader 1 & 4 (3.4\%) & 24 (20.3\%) & 56 (47.5\%) & 34 (28.8\%) & 118 (100\%) \\
Real & Reader 2 & 0 (0\%) & 5 (4.2\%) & 47 (39.8\%) & 66 (55.9\%) & 118 (100\%) \\
Syn. & Reader 1 & 9 (6.9\%) & 22 (16.8\%) & 77 (58.8\%) & 23 (17.6\%) & 118 (100\%) \\
Syn. & Reader 2 & 2 (1.5\%) & 26 (19.8\%) & 89 (67.9\%) & 14 (10.7\%) & 118 (100\%) \\ \bottomrule
\end{tabular}

\smallskip
\footnotesize
\begin{flushleft}
\textbf *Real = Ground truth nephrographic images, *Syn. = Synthesized nephrographic phase images by the ResNCT model
\end{flushleft}
\end{table}

\begin{figure*}[t!]
\centering
\includegraphics[width=1\textwidth]{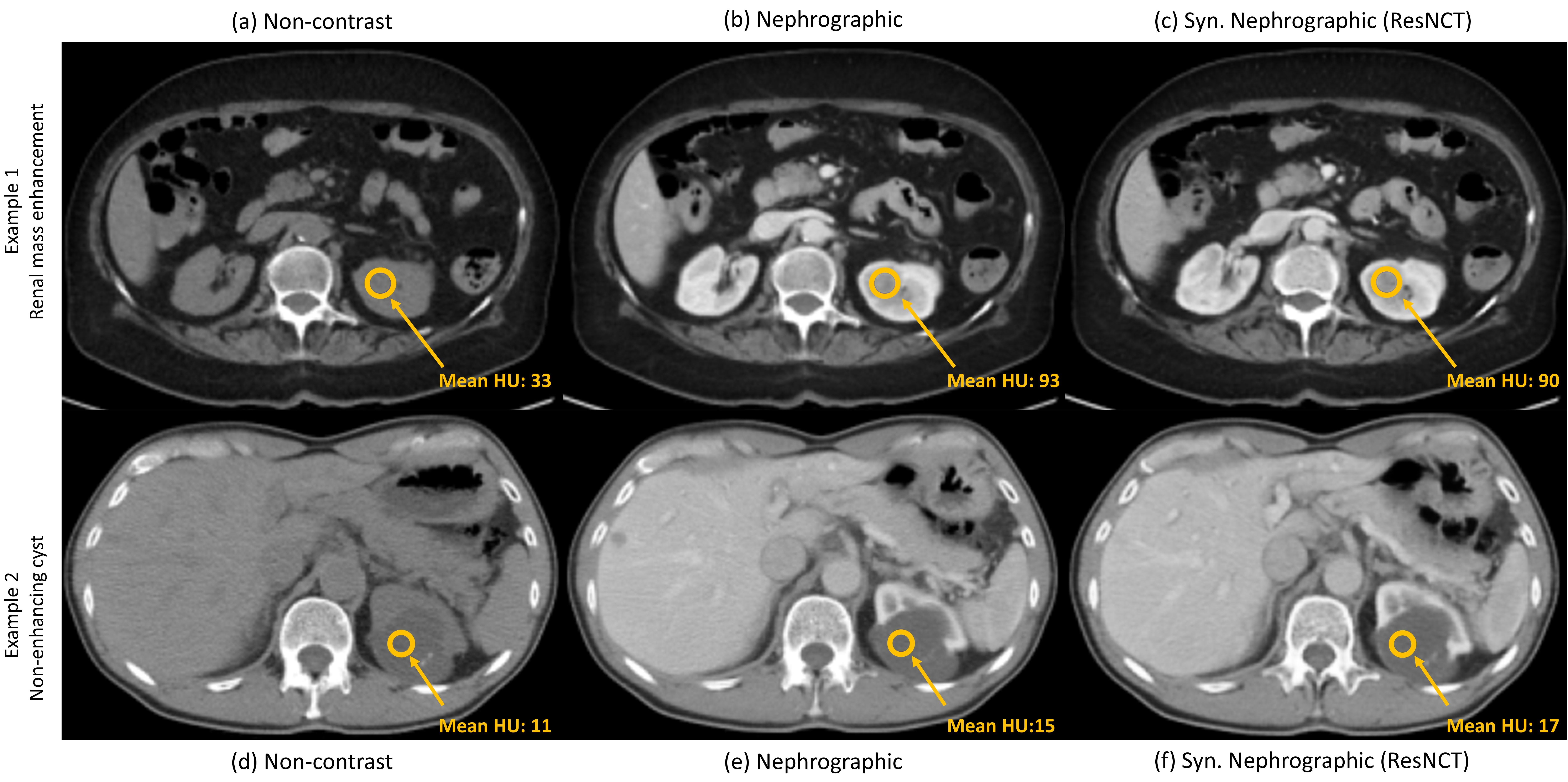}
\caption{\label{fig:fig5.jpg} Attenuation values of an enhancing solid renal mass and a benign simple renal cyst on non-contrast and nephrographic CT urography images, as well as synthesized nephrographic phase images from the ResNCT model. In Example 1 (a-c), a solid renal mass is present in the left kidney, with a mean attenuation of 33 HU on non-contrast images (a). On the nephrographic phase, the attenuation of the mass increases to a mean of 93 HU (b). Similarly, the mean attenuation of the mass was 90 HU on the synthesized nephrographic phase image from the ResNCT model (c). In Example 2, a cystic renal lesion is present in the left kidney, with a mean attenuation of 11 HU on non-contrast images (d). On the nephrographic phase, the attenuation of the lesion is effectively unchanged, with a mean attenuation of 15 HU (e). Similarly, the mean attenuation of the lesion is 17 HU on the synthesized nephrographic phase image from the ResNCT model (f).}
\end{figure*}

\subsection{Image Quality Analysis}
\paragraph{}Two readers provided Likert scores of image quality for ground truth and synthetic nephrographic phase images from the ResNCT model (Table 2). The two readers yielded similar score distributions for synthetic images, but reader 1 tended to assign lower scores to ground truth images than reader 2. No significant difference was found in the scores for Reader 1 between ground truth and synthesized images, with mean scores of 3 ± 0.8 and 2.9 ± 0.8, respectively (p = 0.16). For Reader 2, scores were significantly higher for ground truth than synthesized images, with mean scores of 3.5 ± 0.6 and 2.9 ± 0.6, respectively (p < 0.001). For the two readers, the ICC for ground truth images was -0.049 (95\% confidence intervals: -0.227, 0.132; p = 0.702) and for synthetic images was 0.339 (95\% confidence intervals: 0.178, 0.481; p < 0.001). The overall agreement between the readers yielded an ICC of 0.201 (95\% confidence intervals: 0.079, 0.317; p < 0.001).
\paragraph{}With low (scores of 1-2) and high (scores of 3-4) grouped together for a dichotomized comparison, the agreement between the two reviewers led to a Cohen’s kappa of -0.077 for ground truth images (p = 0.202), 0.191 for synthesized images (p = 0.028), and an overall Cohen’s kappa of 0.083 (p = 0.162).

\section{Discussion}
\paragraph{}This study utilized a deep learning transformer model to reconstruct the nephrographic phase images in CT urography studies from the texture and spatial details obtained from the corresponding non-contrast and urographic CTU phases. This approach enables the reduction of a three-phase acquisition to a two-phase acquisition, effectively lowering the radiation dose by one third compared to the original method.
\paragraph{}Medical synthesis aims to predict the target images from a source under constrained conditions for a given subject. Image synthesis is an ill-posed inverse problem as inferences are made in the absence of a target \cite{ye2013modality}. Deep learning data-driven approaches have helped to overcome this ill-posed problem by learning missing representations directly from data during the model training process. Recent studies have utilized deep learning convolutional networks to synthesis medical images from cross modality domains to predict the target class \cite{huo2018adversarial,huang2017cross,yi2019generative,litjens2017survey}. For example, prior studies \cite{choi2021generating,ristea2023cytran} have employed generative adversarial networks to generate contrast-enhanced images from non-contrast CT images. However, depending on the application, the creation of contrast-enhanced images from non-contrast images is problematic, as the enhancement of a neoplasm is indeterminate on such synthetic images. The methodology utilized in the present study employs both a non-contrast and a contrast-enhanced (urographic) phase to synthesize the target post-contrast images (nephrographic phase), which aids the model in converging to a realistic output.	
\paragraph{}The experimental results suggest that the ResNCT model effectively reconstructed nephrographic phase CT images by incorporating local and global texture as well as spatial information from the non-contrast and urographic phases. Moreover, the synthesized images retained attenuation information with minimal differences between ground truth and synthetic images—a feature not previously included in prior studies of CT image synthesis.
\paragraph{}Likert scale scoring of image quality was performed to compare nephrographic phase images generated from the ResNCT model to the ground truth images. For Reader 1, no significant difference was found in the scores between the synthetic and ground truth images. However, scores were significantly higher for ground truth images than synthetic images for Reader 2. Misregistration between the three CT urography phases was the most likely source of error and artifact in the synthesized images and was most apparent in the bowel. Artifact in the bowel may have led to the generally lower scores in the synthesized images for Reader 2. Future work into improved image registration, as well as further developments in the ResNCT model as discussed below, may mitigate these errors and lead to better image quality scores.
\paragraph{}It is important to note that the current evaluation of the model was based on a strict window level criterion (window of 400, level of 50) to ensure the preservation of values as the images are passed through the ResNCT model. Therefore, voxels with attenuation less than -150 and greater than 250 Hounsfield units (HU) are not well represented on the synthesized images to date. Additional model development is needed to incorporate different window level settings and more data from other CT acquisition systems, presenting an avenue for future exploration to enhance its generalizability. Nevertheless, for soft tissues, the synthesized nephrographic phase images are quantitatively and qualitatively similar to ground truth nephrographic phase images, with minimal attenuation differences between the images as well as the ability to distinguish a non-enhancing structure in the kidney from an enhancing structure. An additional benefit of the synthesized images is the ability to standardize the appearance of the nephrographic phase. Depending on a patient’s renal function and the timing of the acquisition, there is inherent variability in the appearance of the kidneys on the nephrographic phase, with varying degrees of corticomedullary differentiation. A deep learning model that synthesizes the images removes this variability given that the images are created from the non-contrast and urographic phases, which are less prone to such variability.
\paragraph{}The overall scan time of CT urography studies is considerably longer than single-phase CT examinations, given that non-contrast images are first acquired, contrast is then injected, and then at least ten additional minutes are needed until the urographic phase can be acquired. The synthesis of the middle—nephrographic—phase does not impart any time benefits. However, it is an area of active investigation to expand on the deep learning model in this study to CT urography examinations acquired with dual-energy or photon counting CT. Virtual non-contrast images can be created inherently from the data in dual-energy or photon counting CT. As a result, the ResNCT model can be retrained using a dual-energy/photon counting urographic phase and the virtual non-contrast images to synthesize the nephrographic phase. This approach would therefore reduce the overall radiation by 66\%, reduce scan time since only the urographic phase is needed, and also provide inherent image registration given that all time points are derived from a single acquisition.
\paragraph{}While the ResNCT method has demonstrated superiority over the aforementioned generative adversarial methods, it is pertinent to acknowledge that a detailed exploration of this comparison is required with other methods. However, it is crucial to highlight that our future focus will involve an examination of ResNCT against diffusion and transformer models incorporating a more extensive window level spectrum and dataset.
\paragraph{}To date, only GANs and transformer models have been employed for the synthesis of nephrographic images in CTU examinations, which is a limitation of the study. Diffusion models may provide additional benefits for this image synthesis task \cite{dayarathna2023deep}. Diffusion models are computationally intensive, particularly for the volumetric dual inputs needed for this project; thus, a transformer model was used as a first pass implementation for this study. Nevertheless, implementation of a diffusion model is an active area of investigation. The transformer architecture utilized in this study performed well, particularly with respective to the other architectures, as the transformer model enhances feature localization through a multi-attention mechanism and leverages GAN architecture for improved image reconstruction with minimal loss.
\paragraph{}In conclusion, the ResNCT model successfully provided a methodology for synthesizing nephrographic phase images from the other phases in a single-bolus CT urography examination, allowing for a 33\% radiation dose reduction. This model establishes a framework that will allow for additional reductions in radiation dose and scan time, as well as the ability to improve the efficiency of other multi-phase CT examinations.

\bibliographystyle{unsrt}  
\bibliography{references}  

\begin{thebibliography}{10}

\bibitem{Silverman2009current}
Stuart~G Silverman, John~R Leyendecker, and E~Stephen Amis~Jr.
\newblock What is the current role of ct urography and mr urography in the evaluation of the urinary tract?
\newblock {\em Radiology}, 250(2):309--323, 2009.

\bibitem{nawfel2004patient}
Richard~D Nawfel, Philip~F Judy, A~Robert Schleipman, and Stuart~G Silverman.
\newblock Patient radiation dose at ct urography and conventional urography.
\newblock {\em Radiology}, 232(1):126--132, 2004.

\bibitem{cheng2019ct}
Karen Cheng, Fiona Cassidy, Lejla Aganovic, Michael Taddonio, and Noushin Vahdat.
\newblock Ct urography: how to optimize the technique.
\newblock {\em Abdominal Radiology}, 44:3786--3799, 2019.

\bibitem{potenta2015ct}
Scott~E Potenta, Robert D'Agostino, Kevan~M Sternberg, Kanayo Tatsumi, and Karina Perusse.
\newblock Ct urography for evaluation of the ureter.
\newblock {\em Radiographics}, 35(3):709--726, 2015.

\bibitem{chow2007split}
Lawrence~C Chow, Sharon~W Kwan, Eric~W Olcott, and Graham Sommer.
\newblock Split-bolus mdct urography with synchronous nephrographic and excretory phase enhancement.
\newblock {\em American Journal of Roentgenology}, 189(2):314--322, 2007.

\bibitem{raman2017upper}
Siva~P Raman and Elliot~K Fishman.
\newblock Upper and lower tract urothelial imaging using computed tomography urography.
\newblock {\em Radiologic Clinics}, 55(2):225--241, 2017.

\bibitem{morrison2021split}
Nicole Morrison, Sherrie Bryden, and Andreu~F Costa.
\newblock Split vs. single bolus ct urography: Comparison of scan time, image quality and radiation dose.
\newblock {\em Tomography}, 7(2):210--218, 2021.

\bibitem{dillman2007comparison}
Jonathan~R Dillman, Elaine~M Caoili, Richard~H Cohan, James~H Ellis, Isaac~R Francis, Bin Nan, and Yong Zhang.
\newblock Comparison of urinary tract distension and opacification using single-bolus 3-phase vs split-bolus 2-phase multidetector row ct urography.
\newblock {\em Journal of computer assisted tomography}, 31(5):750--757, 2007.

\bibitem{pasternak2012clinical}
Jeffrey~J Pasternak and Eric~E Williamson.
\newblock Clinical pharmacology, uses, and adverse reactions of iodinated contrast agents: a primer for the non-radiologist.
\newblock In {\em Mayo Clinic Proceedings}, volume~87, pages 390--402. Elsevier, 2012.

\bibitem{dalmaz2022resvit}
Onat Dalmaz, Mahmut Yurt, and Tolga {\c{C}}ukur.
\newblock Resvit: residual vision transformers for multimodal medical image synthesis.
\newblock {\em IEEE Transactions on Medical Imaging}, 41(10):2598--2614, 2022.

\bibitem{schwartz1994kidney}
Laurent~H Schwartz, Jocelyne Richaud, Laurent Buffat, Emmanuel Touboul, and Michel Schlienger.
\newblock Kidney mobility during respiration.
\newblock {\em Radiotherapy and Oncology}, 32(1):84--86, 1994.

\bibitem{isola2017image}
Phillip Isola, Jun-Yan Zhu, Tinghui Zhou, and Alexei~A Efros.
\newblock Image-to-image translation with conditional adversarial networks.
\newblock In {\em Proceedings of the IEEE conference on computer vision and pattern recognition}, pages 1125--1134, 2017.

\bibitem{saxena2021comparison}
Sagar Saxena and Mohammad~Nayeem Teli.
\newblock Comparison and analysis of image-to-image generative adversarial networks: a survey.
\newblock {\em arXiv preprint arXiv:2112.12625}, 2021.

\bibitem{ye2013modality}
Dong~Hye Ye, Darko Zikic, Ben Glocker, Antonio Criminisi, and Ender Konukoglu.
\newblock Modality propagation: coherent synthesis of subject-specific scans with data-driven regularization.
\newblock In {\em Medical Image Computing and Computer-Assisted Intervention--MICCAI 2013: 16th International Conference, Nagoya, Japan, September 22-26, 2013, Proceedings, Part I 16}, pages 606--613. Springer, 2013.

\bibitem{huo2018adversarial}
Yuankai Huo, Zhoubing Xu, Shunxing Bao, Albert Assad, Richard~G Abramson, and Bennett~A Landman.
\newblock Adversarial synthesis learning enables segmentation without target modality ground truth.
\newblock In {\em 2018 IEEE 15th international symposium on biomedical imaging (ISBI 2018)}, pages 1217--1220. IEEE, 2018.

\bibitem{huang2017cross}
Yawen Huang, Ling Shao, and Alejandro~F Frangi.
\newblock Cross-modality image synthesis via weakly coupled and geometry co-regularized joint dictionary learning.
\newblock {\em IEEE transactions on medical imaging}, 37(3):815--827, 2017.

\bibitem{yi2019generative}
Xin Yi, Ekta Walia, and Paul Babyn.
\newblock Generative adversarial network in medical imaging: A review.
\newblock {\em Medical image analysis}, 58:101552, 2019.

\bibitem{litjens2017survey}
Geert Litjens, Thijs Kooi, Babak~Ehteshami Bejnordi, Arnaud Arindra~Adiyoso Setio, Francesco Ciompi, Mohsen Ghafoorian, Jeroen~Awm Van Der~Laak, Bram Van~Ginneken, and Clara~I S{\'a}nchez.
\newblock A survey on deep learning in medical image analysis.
\newblock {\em Medical image analysis}, 42:60--88, 2017.

\bibitem{choi2021generating}
Jae~Won Choi, Yeon~Jin Cho, Ji~Young Ha, Seul~Bi Lee, Seunghyun Lee, Young~Hun Choi, Jung-Eun Cheon, and Woo~Sun Kim.
\newblock Generating synthetic contrast enhancement from non-contrast chest computed tomography using a generative adversarial network.
\newblock {\em Scientific reports}, 11(1):20403, 2021.

\bibitem{ristea2023cytran}
Nicolae-C{\u{a}}t{\u{a}}lin Ristea, Andreea-Iuliana Miron, Olivian Savencu, Mariana-Iuliana Georgescu, Nicolae Verga, Fahad~Shahbaz Khan, and Radu~Tudor Ionescu.
\newblock Cytran: a cycle-consistent transformer with multi-level consistency for non-contrast to contrast ct translation.
\newblock {\em Neurocomputing}, 538:126211, 2023.

\bibitem{dayarathna2023deep}
Sanuwani Dayarathna, Kh~Tohidul Islam, Sergio Uribe, Guang Yang, Munawar Hayat, and Zhaolin Chen.
\newblock Deep learning based synthesis of mri, ct and pet: Review and analysis.
\newblock {\em Medical Image Analysis}, page 103046, 2023.

\end{thebibliography}

\end{document}